\documentstyle[12pt]{article}

\begin{document}

\begin{center}
{\bf Condition on the symmetry-breaking solution of the Schwinger-Dyson
Equation}

G. Cheng\\[0pt]
CCAST (World Laboratory), P.O. Box 8730, Beijing 100080, China\\[0pt]
and\\[0pt]
Fundamental Physics Center, Univ. of Sci. and Tech. of China,\\[0pt]
Heifei, Anhui, 230026, P.R. China$^*$\\[0pt]
$~$\\[0pt]
T.K. Kuo\\[0pt]
Physics Department, Purdue University, West Lafayette, Indiana 47907\\[0pt]
\end{center}

\vspace{10pt}

\begin{center}
{\bf Abstract}
\end{center}

\begin{quote}
$~~~~~$We derive a condition for a non-trivial solution of the
Schwinger-Dyson equation to be accompanied by a Goldstone bound state. It
implies that, for quenched planar QED, although chiral symmetry breaking
occurs when there is a cutoff, the continuum limit fails to exist.
\end{quote}

\vspace{40pt}

\noindent $^*$Mailing address.

\pagebreak

\noindent {\bf I. Introduction}

Dynamical chiral symmetry breaking$^1$ (DCSB) has been extremely useful in a
wide variety of topics in physics, from the success of low energy theorems
in current algebra, to the construction of composite models of the Higgs
boson. There are two aspects in the mechanism of DCSB, the mass generation
of fermions and the existence of the Goldstone boson.$^2$ More concretely,
chiral symmetry breaking occurs when the self-consistent Schwinger-Dyson
(SD) equation for the fermion mass develops a non-trivial solution as the
coupling constant reaches a critical value.$^3$ In addition, for the same
coupling constant, there must also exist a bound state solution of the
Bethe-Salpeter (BS) equation, with vanishing 4-momenta, corresponding to the
massless pseudoscalar Goldstone boson. When both of these conditions are
met, the self-energy solution shall be called a symmetry-breaking solution.
We should also mention that T. Maskawa and H. Nakajima$^4$ had proposed an
equivalent condition for the existence of the symmetry-breaking solution.
Their condition is that the self-energy solution has to satisfy the
Ward-Takahashi identity.

For definiteness, we will concentrate on QED in the massless, quenched
planar approximation$^5$, which is a well-known model in the study of DCSB.
It is simple enough for a detailed study on its properties, yet it is also
rich in structure which suggests general physical features in other models.
Thus, with the incorporation of running coupling constant, it can be easily
extended to QCD in the large N approximation.$^6$ In addition, results
gleaned from studies of this model have been used in the construction of
realistic models in the context of technicolor$^7$, $t\bar{t}$ condensation$%
^8$, etc.

In the study of the quenched planar QED, an important question concerns the
cutoff. Thus, Maskawa and Nakajima established the existence of a critical
coupling constant $\alpha_c$ in the solution of the SD equation with a
cutoff, $\Lambda$. When $\alpha > \alpha_c$, a nontrivial solution of the SD
equation emerges which signals the advent of chiral symmetry breaking. A
natural question is whether there exists a limit, $\alpha_c(\infty) =
\lim_{\Lambda \rightarrow \infty}\alpha_c(\Lambda)$, corresponding to the
existence of a spontaneously broken phase in the continuum limit.$^3$ That
such an ultraviolet fixed point exists is often accepted without close
examinations. In this work we will study this important question in detail.
To this end we consider the BS equation with the quantum numbers $\ell^P =
0^-$ and vanishing 4-momentum. The mass in the fermion propagator contained
in the BS equation is identified with the self-mass of the non-trivial
solution of the SD-equation. A symmetry-breaking solution of the SD equation
must be one which is associated with a pseudoscalar bound state solution of
the BS equation.

It can be established that the bound state wave function is related to the
self-energy function. Thus, the normalization condition of the bound state
wave function is converted into one for the self-energy function. In this
way, we find a necessary and sufficient condition for the non-trivial
solution of the SD equation to be symmetry-breaking. When we apply this
result to the quench planar QED, we find that the critical point does not
have a limit. Thus, the ultraviolet fixed point can not exist in quenched
planar QED in the continuum limit, although DCSB is established in the case
with a cutoff. \vspace{10pt}

\noindent {\bf II. Schwinger-Dyson and Bethe-Salpeter Equations} \vspace{10pt%
}

>From the Goldstone theorem, it is well-known that a massless boson appears
whenever a generator of a global continuous symmetry group is broken
spontaneously. The quantum number of this particle corresponds to that of
the operator of the current which is broken. For massless quenched QED, the
broken current is the electromagnetic axial current so that the Goldstone
particle is a pseudoscalar boson. We associate this to the bound state
solution of the BS-equation, in which the fermion mass will be identified
with the dynamically generated mass from the SD equation. This approach has
a number of advantages in quenched planar QED. First, in the ladder
approximation, unitarity in elastic scattering is restored and the problem
of abnormal states is alleviated.$^9$ Second, the approximation in the SD
and BS equations are self-consistent. Last, but not least, the requirement
that the Goldstone boson accompany the fermion mass generation in ladder QED
is satisfied automatically when the condition for a symmetry-breaking
solution is met. In the following our discussions will be in this framework
and we will deduce the condition for the SD self-energy solution to be
symmetry-breaking.

The renormalized fermion propagator will be written as 
$$
S_f^{-1}(p) = \gamma\cdot p~\alpha(p^2) - \beta(p^2) \eqno{(1)} 
$$
The SD equation for the fermion self-energy is: 
$$
S_f^{-1}(p) = S_f^{0-1}(p) + \frac{ie^2}{(2\pi)^4} \int d^4 qD^{\mu\nu}(p-q)
\Gamma_\mu(p,q)S_f(q)\gamma_\nu \eqno{(2)} 
$$

In the quenched planar approximation, we have 
$$
D^{\mu\nu}(p-q) \approx \frac{1}{(p-q)^2} \left[ -g^{\mu\nu} + (1-\xi) \frac{%
(p-q)^\mu(p-q)^\nu}{(p-q)^2}\right] \eqno{(3)} 
$$
and 
$$
\Gamma_\mu(p,q) \approx \gamma_\mu, \eqno{(4)} 
$$
where $\xi$ is a gauge parameter, with $\xi = 0$ and $\xi = 1$ corresponding
to the Landau and Feynman gauges, respectively. Substituting Eqs.(3),(4) and
(1) into (2), after the Wick rotation and taking trace on both sides, we
obtain: 
$$
\beta(p^2) = \frac{e^2}{(2\pi)^4} \int d^4q \frac{4-(1-\xi)}{(p-q)^2} \frac{%
\beta(q^2)}{q^2\alpha^2(q^2)+\beta^2(q^2)} \eqno{(5)} 
$$
Similarly, with the multiplication of $\gamma\cdot p$, we have: 
$$
\everymath={\displaystyle} 
\begin{array}{rcl}
\alpha(p^2) & = & 1 + \frac{e^2}{(2\pi)^4} \frac{1}{p^2} \int d^4q \frac{1}{%
(p-q)^2}\times \frac{\alpha(q^2)}{q^2\alpha^2(q^2)+\beta^2(q^2)} \times \\ 
&  &  \\ 
&  & \times \left[ 2p\cdot q+(1-\xi)\left\{ \frac{2[p\cdot(p-q)][q\cdot(p-q)]%
}{(p-q)^2} - p\cdot q\right\}\right] .
\end{array}
\eqno{(6)} 
$$

Eqs.(5) and (6) determine the self-energy of the fermion in massless
quenched QED. The trivial solution, $\beta(p^2)=0$ and $\alpha(p^2)$ =
finite ( = 1, in the Landau gauge), corresponds to that of the symmetric
vacuum. But it is certainly true that not all of the nontrivial solutions
signal spontaneous symmetry breaking. The criterion that a nontrivial $%
\beta(p^2)$ does trigger DCSB is that there must be an accompanying
Goldstone boson. That is to say, there must be a bound state solution of the
BS equation, with the appropriate quantum numbers and at the same coupling
constant.

For a bound state composed of the fermion anti-fermion pair $(A$ and $\bar{B}
$), in the ladder approximation in QED, the wave function is determined by:$%
^{10}$%
$$
S_{f}^{A^{-1}}(\frac{k}{2}+p)\chi _{k}(p)\lbrack S_{f}^{B^{-1}}(\frac{k}{2}%
-p)=\frac{ie^{2}}{(2\pi )^{4}}~\int ~d^{4}qD^{\mu \nu }(p-q)\Gamma _{\mu
}^{A}(p,q)\chi _{k}(q)\gamma _{v}^{B}\eqno{(7)}
$$

$$
\everymath={\displaystyle}
\begin{array}{rcl}
&  & \lbrack (\frac{k}{2}+p)\cdot \gamma \alpha (p^{2})-\beta (p^{2})\rbrack
\chi _{k}(p)\lbrack (\frac{k}{2}-p)\cdot \gamma \alpha (p^{2})+\beta
(p^{2})\rbrack  \\ 
& = & \frac{ie^{2}}{(2\pi )^{4}}~\int ~d^{4}qD^{\mu \nu }(p-q)\Gamma _{\mu
}(p,q)\chi _{k}(q)\gamma _{v}
\end{array}
\eqno{(8)}
$$

\noindent Taking the same approximation as in the case of the SD equation
and putting Eqs. (3) and (4) into Eq. (18), the BS equation reads:

$$
\everymath={\displaystyle} 
\begin{array}{rcl}
&  & [(\frac{k}{2} + p ) \cdot \gamma\alpha (p^2) - \beta(p^2)] \chi_k(p) [(%
\frac{k}{2} - p) \cdot \gamma\alpha (p) + \beta(p^2)] = \\ 
&  & = \frac{ie^2}{(2\pi)^4} \int d^4q \frac{1}{(p-q)^2} [ -g^{\mu\nu} +
(1-\xi) \frac{(p-q)^\mu(p-q)^\nu}{(p-q)^2}] \gamma_\mu\chi_k(q)\gamma_\nu
\end{array}
\eqno{(9)} 
$$
This is the form of the BS equation in quenched planar QED.

Since the Goldstone state has $\ell^P =0^-$ and vanishing mass, the wave
function $\chi_k(p)$ is spherically symmetric. Taking $k=0$, we write 
$$
\chi_k(p) = \chi_0^P(p^2)\times \gamma_5 \eqno{(10)} 
$$
Substituting Eq.(10) into Eq.(9) and taking trace after Wick rotation, we
obtain 
the BS equation for the Goldstone boson: 
$$
\left[\alpha^2(p^2)p^2 + \beta^2(p^2)\right]  \chi_0^P(p^2) = \frac{e^2}{%
(2\pi)^4}  \int d^4q \frac{4-(1-\xi)}{(p-q)^2}  \chi_0^P(q^2) \eqno{(11)} 
$$
\vspace{10pt}

\noindent {\bf III. Condition for a symmetry-breaking solution of the SD
equation} \vspace{10pt}

For any physical bound state, the wave function must satisfy a normalization
condition. For the Goldstone boson wave function, Eq.(11), there were
several approaches with equivalent results.$^{11}$ We shall quote the
normalization condition in the following form,$^{12}$ which is valid for a
fermion-antifermion system in ladder QED, 
$$
\everymath={\displaystyle} 
\begin{array}{rcl}
&  & \int d^4 qT_r \biggl\{ \bar{\chi}_k(q)\left[(\frac{k}{2} +q) \cdot
\gamma \alpha(q^2) - \beta(q^2)\right] \chi_k(q) \times \\ 
&  & \times \left[(\frac{k}{2}-q)\cdot \gamma\alpha(q^2) + \beta(q^2)\right]%
\biggl\} = \lambda \frac{dM_B}{d\lambda}
\end{array}
\eqno{(12)} 
$$
Here, $M_B$ is the mass of the bound state and $\lambda$ is the coupling
constant, $\lambda = \alpha/4\pi$.

For a bound state with vanishing mass, vanishing total angular momentum and
negative parity, eq.(12) become: 
$$
\int d^4q [\alpha^2(q^2)q^2 + \beta^2(q^2)]  |\chi_0^P(q^2)|^2 = ~finite %
\eqno{(13)} 
$$
This is the condition for $\chi$ to be the wave function of a physical
Goldstone boson which accompanies the non-trivial solution of the SD
equation.

Due to the spherical symmetry, we can integrate Eqs.(11), (13),(5), and (6)
with respect to the angular variables in the 4-dimensional Euclidean space.
They are then simplified into the following set of equations: 
$$
[p^2\alpha^2(p^2) + \beta^2(p^2)] \chi_0^P(p^2)  = \frac{\alpha}{4\pi^3}
\int_\epsilon^\Lambda~  dq^2K_\beta(p^2,q^2)q^2\chi_0^P(q^2) \eqno{(14)} 
$$

$$
\int_\epsilon^\Lambda dq^2 q^2[\alpha^2(q^2)q^2 
+\beta^2(q^2)]|\chi_0^P(q^2)|^2 =~finite \eqno{(15)} 
$$

$$
\beta(p^2) = \frac{\alpha}{4\pi^3} \int_\epsilon^\Lambda 
dq^2K_\beta(p^2,q^2) \frac{q^2\beta(q^2)} {q^2\alpha^2(q^2)+\beta^2(q^2)} %
\eqno{(16)} 
$$

$$
\alpha(p^2) =1 + \frac{\alpha}{4\pi^3} \int_\epsilon^\Lambda 
dq^2K_\alpha(p^2,q^2) \frac{q^2\alpha(q^2)} {q^2\alpha^2(q^2)+\beta^2(q^2)} %
\eqno{(17)} 
$$

The wave function $\chi_0^P(p^2)$ is determined by Eq.(14), with the
normalization condition Eq.(15). Eqs.(16) and (17), which determine the
self-energy functions $\alpha(p^2)$ and $\beta(p^2)$, have been written with
explicit ultraviolet and infrared cutoffs. The continuum limit corresponds
to $\Lambda \rightarrow \infty$ and $\epsilon \rightarrow 0$. In these
equations, we have also defined $\alpha = e^2/4\pi$. Lastly, the kernels $%
K_\alpha$ and $K_\beta$ are given by:

$$
\everymath={\displaystyle} 
\begin{array}{rcl}
K_\alpha(p^2,q^2) & = & \frac{2\pi}{p^2}\int_o^\pi~ d\omega \sin^2 \omega 
\frac{1}{(p-q)^2} \times \\ 
&  &  \\ 
& \times & \left[ 2p\cdot q + (1-\xi) \left( \frac{2p\cdot(p-q)q\cdot(p-q)} {%
(p-q)^2}- p\cdot q \right) \right],
\end{array}
\eqno{(18)} 
$$

$$
\everymath={\displaystyle} 
\begin{array}{rcl}
K_\beta(p^2,q^2) & = & 2\pi \int_o^\pi d\omega \sin^2 \omega \frac{4-(1-\xi)%
}{(p-q)^2} \\ 
& = & (3+\xi)\pi^2 \left[ \frac{\theta(p^2-q^2)}{p^2} + \frac{\theta(q^2-p^2)%
}{q^2} \right].
\end{array}
\eqno{(19)} 
$$
If we introduce the function 
$$
\psi(p^2) = [p^2\alpha(p^2)+\beta^2(p^2)]  \chi_0^P(p^2) \eqno{(20)} 
$$
then eq.(14) becomes: 
$$
\psi(p^2) = \frac{\alpha}{4\pi^3} \int_\epsilon^\Lambda  dq^2
K_\beta(p^2,q^2) \frac{q^2\psi(q^2)} {q^2\alpha^2(q^2)+\beta^2(q^2)}. %
\eqno{(21)} 
$$
Also the normalization condition reads: 
$$
\int_\epsilon^\Lambda dq^2 \frac{q^2}{q^2\alpha^2(q^2) + \beta^2(q^2)} 
|\psi(q^2)|^2 = ~finite. \eqno{(22)} 
$$
In order for spontaneous symmetry breaking to occur, Eq.(16) must have a
non-trivial solution, $\beta(p^2) \not{= }0$. However, this is a necessary,
but not a sufficient condition. For $\beta(p^2)$ to be symmetry-breaking, it
must satisfy an additional normalization condition. For clarity of
presentation, we now state this result in the form of the following theorem.

{\bf Theorem:} The necessary and sufficient condition for a non-vanishing
solution $\beta(p^2)$ of Eq.(16) to be symmetry-breaking is that it must
satisfy, together with the solution $\alpha(p^2)$ of Eq.(17), the condition: 
$$
\int_\epsilon^\Lambda dq^2 \frac{q^2}{q^2\alpha^2(q^2)+\beta^2(q^2)} 
|\beta(q^2)|^2 = ~finite. \eqno{(23)} 
$$

{\bf Proof:} If, for some coupling constant $\alpha = \alpha_i$, there are
non-vanishing solutions of Eqs.(16) and (17), $\alpha(p^2)$ and $\beta(p^2)$%
, which satisfy the condition Eq.(23), then we have a solution of Eq.(21),
with 
$$
\psi(p^2) = c\beta(p^2) \eqno{(24)} 
$$
for an arbitrary constant $c$. It also satisfies Eq.(22) and is thus a bound
state solution. This proves that Eq.(23) is a sufficient condition. To show
that it is also necessary, we need to make sure that the Goldstone boson
wave function is nondegenerate, so that the solution in Eq.(24) is unique.
The nondegeneracy of the Goldstone mode follows from the fact that it comes
from the breaking of the generator of an abelian group.$^1$ This unique
solution must satisfy the normalization condition Eq.(15), which, in turn,
implies Eq.(22). This means that for $\beta(p^2)$ to be symmetry-breaking,
it must satisfy Eq.(23). \vspace{10pt}

\noindent {\bf IV. An Application} \vspace{10pt}

The results obtained above will now be applied to QED in the quenched planar
approximation. In the Landau gauge, $\xi = 0$. Then Eq. (18) yields$^{(13)}$ 
$K_\alpha = 0$ and $\alpha(p^2) = 1$. This is a property of the quenched
planar approximation so that there is no wave function renormalization, $Z_2
= 1$, implying $Z_1 = 1$, from $Z_1 = Z_2$. also, in the quenched
approximation there are no photon self-energy and coupling constant
renormalizations, $Z_3 = 1$. Thus, within this model, renormalization does
not arise. It is obviously interesting and important to address this
problem. However, it is beyond the scope of the present work.

Let us now concentrate on the function $\beta(p^2)$, with $\alpha(p^2) = 1$
and using Eq. (19), Eq. (16) reads: 
$$
\beta(p^2) = \frac{3\alpha}{4\pi} \int_\epsilon^\Lambda  dq^2 \left[ \frac{%
\theta(p^2-q^2)}{p^2}  + \frac{\theta(q^2-p^2)}{q^2}\right] \frac{%
q^2\beta(q^2)}{q^2+\beta^2(q^2)} \eqno{(25)} 
$$
Even for this simplified equation, one does not have an analytic solution.
However, for the purpose of determining the points of phase transition, we
can make use of the bifurcation theory. It states that nontrivial solutions
of nonlinear equations such as Eq.(25) branch off from the trivial solution
at critical values of the coupling constant, $\alpha = \alpha_i$. Such
bifurcation points correspond to phase transition points and can be found by
linearization. That is, taking Frechet derivative or, equivalently, making
an expansion around $\beta = 0$. The Frechet derivative of Eq.(25) is 
$$
d\beta(p^2) = \frac{3\alpha}{4\pi}  \int_\epsilon^\Lambda \left[\frac{%
\theta(p^2-q^2)}{p^2}  + \frac{\theta(q^2-p^2)}{q^2} \right]  d\beta(q^2). %
\eqno{(26)} 
$$

To look for branch points of Eq.(25), there are two useful theorems in
bifurcation theory. 1. For $\alpha_i$ to be a critical point of Eq.(25), it
must be an eigenvalue of Eq.(26). 2. If the operator of the Frechet
derivative is compact and if $\alpha_i \not{= }0$ is one of its eigenvalues
with odd multiplicity, then $\alpha_i$ is a branch point of Eq.(25). Thus,
to find the bifurcation points of Eq.(25), one need only to solve the
eigenvalue problem of Eq.(25), for which general solutions were studied by a
number of authors. In particular, converting it into a differential
equation, Kondo et al.$^{14}$ found the solution:

\[
\everymath={\displaystyle}  ~~~~~\beta(p^2) = \Biggl\{ 
\begin{array}{lcr}
Ap^{-1+2\sigma} + Bp^{-1-2\sigma}, & 0 < \lambda < \lambda_c, & 
~~~~~~~~~~~~~~~~~~~~~~~~~~~ (27) \\ 
p^{-1}(C+D\ell np^2), & \lambda = \lambda_c & ~~~~~~~~~~~~~~~~~~~~~~~~~~~
(28) \\ 
Ep^{-1+i2\rho} + Fp^{-1-i2\rho}, & \lambda_c < \lambda & 
~~~~~~~~~~~~~~~~~~~~~~~~~~~ (29)
\end{array}
\]
where 
\[
\sigma = \frac{1}{2} \sqrt{1 - \frac{\lambda}{\lambda_c}},  \rho = \frac{1}{2%
} \sqrt{\frac{\lambda}{\lambda_c} -1},  ~~~~~~~  \lambda = \frac{\alpha}{4\pi%
}, \alpha_c = \frac{\pi}{3}, 
\]
They will have to satisfy boundary conditions at $p^2 = \epsilon$ and $p^2 =
\Lambda$, which are determined by the integral equation, Eq.(25). These
conditions determine the coefficients $A,B,...$ up to a multiplicative
constant together with the constraints: 
$$
\left[ (\frac{1}{2} + \sigma)^2 - (\frac{1}{2} - \sigma)^2  (\frac{\epsilon}{%
\Lambda})^{2\sigma}\right]  / \left[1 - (\frac{\epsilon}{\Lambda}%
)^{2\sigma}\right] = 0,  ~~~~~~~~~~~  0 < \lambda < \lambda_c; \eqno{(30)} 
$$

$$
\frac{1}{4} \left\{ 1 + \frac{4}{[\ell n(\Lambda/\epsilon)]} \right\} = 0,
~~~~~~~~  \lambda = \lambda_c; \eqno{(31)} 
$$

$$
\left(\left\{ n\pi + arctan\left[\rho/(\rho^2-\frac{1}{4}  )\right] \right\}
/\rho\right) = \ell n \frac{\Lambda}{\epsilon},  ~~~~~~~~~~~~ \lambda_c <
\lambda . \eqno{(32)} 
$$

In the physically interesting situation, when $\epsilon/\Lambda$ is small it
is seen that Eqs.(30) and (31) can never be satisfied. However, Eq.(32) has
a sequence of solutions, $\lambda = \lambda_i$. They are dependent on $%
\epsilon\Lambda$. Also it can be shown that these solutions are
nondegenerate. Thus, for quenched planar QED with given $\epsilon$ and $%
\Lambda$, the solutions $\lambda = \lambda_i$ signal the onset of dynamical
symmetry breaking.

But this conclusion is not valid in the continuum limit, $\epsilon
\rightarrow 0$ and $\Lambda \rightarrow \infty$. Here, one needs to examine
the normalization condition, Eq.(23), that the solution $\beta(p^2)$ must
satisfy. A straightforward calculation shows that the solutions obtained
above do not satisfy Eq.(23), when $\epsilon \rightarrow 0$ and $\Lambda
\rightarrow \infty$. We conclude that actually there does not exist a value
of $\lambda$ for which quenched planar QED, in the continuum limit,
undergoes dynamical symmetry breaking.

\noindent {\bf V. Conclusions} \vspace{10pt}

Nontrivial solutions of the self-energy SD equation have long been
associated with dynamical symmetry breaking. However, true DCSB also
requires the existence of the Goldstone boson. We will take this to mean
that the non-trivial solution of the SD equation must be associated with a
solution of the BS equation which has appropriate quantum numbers and zero
4-momentum. In this case, the solution of the SD equation will be called
symmetry-breaking, which is the condition for the occurrence of DCSB.

In this work, we studied in detail the symmetry-breaking solution of the SD
equation in quenched planar QED. Chiral symmetry breaking occurs when a
nontrivial solution of SD equation is associated with a $\ell^P = 0^-$
Goldstone bound state solution of the BS equation. It turns out that the
self-energy function of the SD equation is related to the bound state wave
function of the BS equation. Thus, the normalization condition on the bound
state wave function turns into a constraint equation on the self-energy
function through Eq.(20). We applied this constraint to the known solutions
of quenched planar QED, and found that, although DCSB occurs when cutoffs
are present, the continuum limit does not exist.

In the study of dynamical symmetry breaking, it is important to have a
criterion for establishing whether a nontrivial solution of the SD equation
does signal symmetry breaking. We have found an explicit condition for
quenched planar QED. We are now investigating the problem for other models
and it should be useful in the general study of phase transitions, such as
was done using the bifurcation theory.$^{15}$ We hope to present these
results in the future. \vspace{10pt}

\noindent {\bf Acknowledgements} \vspace{10pt}

G.C. would like to express his sincere thanks to Profs. M.L. Yan, Z.X.
Zhang, G.D. Zhao, H.A. Peng and Y.P. Kuang for their interest and fruitful
discussions. G.C. is supported in part by the National Science Foundation
and the NDSTPR Foundation in China. T.K. is supported by a DOE grant,
DE-FG02-91ER40681.

\pagebreak

\begin{center}
{\bf References}
\end{center}

\begin{enumerate}

\item  {\it Dynamical Gauge Symmetry Breaking}, ed. E. Farhi and R. Jackiw,
Word Scientific, (1982); {\it Proceedings of the 1991 Nagoya Spring School
on Dynamical Symmetry Breaking}, ed. K. Yamawaki, Word Scientific,
Singapore, (1992).

\item  J. Goldstone, {\it Nuovo Cimento}, {\bf 19}, 154 (1961). 

\item  P.I. Fomin, V.P. Gusynin, V.A. Miransky, and Yu. A. Sitenko, {\it %
Riv. Nuovo Cim. Soc. Ital. Fis.} {\bf 6}, 1 (1983); C.N. Leung, S.T. Love,
and W.A. Bardeen, {\it Nucl. Phys.} {\bf B273}, 649 (1986); {\bf B323}, 493
(1989). 

\item  T. Maskawa and H. Nakajima, {\it Prog. Theor. Phys.} {\bf 52}, 1326
(1974); {\bf 54}, 860 (1975). 

\item  K. Johnson, M. Baker, and R. Wiley, {\it Phys. Rev.} {\bf 136}, B1111
(1964); {\bf 163}, 1699 (1967); S.L. Adler and W.A. Bardeen, {\it Phys. Rev.}
{\bf D4}, 3045 (1971). 

\item  K. Higashijima, {\it Phys. Rev.} {\bf D29}, 1228 (1986). 

\item  T. Appelquist, D. KarabalI, L.C.R. Wijewardhana, {\it Phys. Rev. Lett.%
} {\bf 57}, 957 (1986); T. Appelquist, D. Carrier, L.C.R. Wijewardhana, and
W. Zheng, {\it Phys. Rev. Lett.} {\b 60}, 1114 (1988); T. Appelquist, K.
Lane, and U. Mahanta, {\it Phys. Rev. Lett.} {\bf 61}, 1553 (1988). 

\item  V.A. Miransky, M. Tanabashi, and K. Yamawaki, {\it Mod. Phys. Lett.} 
{\bf A4}, 1043 (1989); {\it Phys. Lett.} {\bf B221}, 177 (1989); W.J.
Marciano, {\it Phys. Rev. Lett.} {\bf 62}, 2793 (1989); W.A. Bardeen, C.T.
Hill, and M. Lindner, {\it Phys. Rev.} {\bf D41}, 1647 (1990).

\item  G. Cheng, {\it Comm. Theor. Phys.} {\bf 15}, 219 (1991); {\bf 15},
303 (1991); G. Cheng and L.Y. Li, {\it Comm. Theor. Phys.} {\bf 15}, 451
(1991); M.J. Levine and J. Wright, {\it Phys. Rev.} {\bf 154}, 1433 (1967); 
{\bf 157}, 1416 (1967). 

\item  N. Nakanishi, {\it Prog. Theor. Phys. Suppl.} {\bf 95}, 1 (1988); 
{\bf 43}, 1 (1969). 

\item  D. Lurie, {\it Particles and Fields}, John Wiley \& Sons, New York
London Sydney, (1968). 

\item  See Reference 10 and L.G. Suttorp, {\it Il Nuovo Cimento}, {\bf 29A},
225 (1975); {\bf 33A}, 257 (1976). 

\item  See, for instance, R. Fukuda and T. Kugo, Nucl. Phys. {\bf B117}, 250
(1976); D. Curtis and M. Pennington, Phys. Rev. {\bf D48}, 4933 (1993). 

\item  K. Kondo, J. Mino, and K. Yamawaki, {\it Phys. Rev.} {\bf D39}, 2430
(1989). 

\item  G. Cheng and T.K. Kuo, {\it J. Math. Phys.} {\bf 35}, 6270 (1994); 
{\bf 35}, 6693 (1994).
\end{enumerate}

\end{document}